\documentclass[twocolumn,aps,showpacs,amsmath,floatfix,superscriptaddress]{revtex4}
\usepackage{graphicx}
\usepackage{dcolumn}
\usepackage{bm}
\usepackage{hyperref} 
\usepackage{latexsym}
\usepackage{float}

\begin{document}
\title{A Billiard-Theoretic Approach to Elementary 1d Elastic Collisions}
\author{S.~Redner}
\email{redner@bu.edu}
\affiliation{Center for BioDynamics, Center for Polymer Studies, 
and Department of Physics, Boston University, Boston, MA, 02215}

\begin{abstract}
  
  A simple relation is developed between elastic collisions of freely-moving
  point particles in one dimension and a corresponding billiard system.  For
  two particles with masses $m_1$ and $m_2$ on the half-line $x>0$ that
  approach an elastic barrier at $x=0$, the corresponding billiard system is
  an infinite wedge.  The collision history of the two particles can be
  easily inferred from the corresponding billiard trajectory.  This
  connection nicely explains the classic demonstrations of the ``dime on the
  superball'' and the ``baseball on the basketball'' that are a staple in
  elementary physics courses.  It is also shown that three elastic particles
  on an infinite line and three particles on a finite ring correspond,
  respectively, to the motion of a billiard ball in an infinite wedge and on
  on a triangular billiard table.  It is shown how to determine the angles of
  these two sets in terms of the particle masses.
  
\end{abstract}
\maketitle

\section{INTRODUCTION}

A standard discussion topic in freshman mechanics courses is elastic
collisions.  There are two very nice lecture demonstrations to accompany this
topic \cite{ped}.  The first is the ``dime on the superball''.  Here one
carefully places a dime on top of a superball and drops this composite system
on a hard floor.  Before doing so, the instructor asks the class to guess the
maximum height of the dime $h_{\rm max}$ compared to the initial height
$h_0$.  For perfectly elastic collisions and in the limit where the mass of
the dime is vanishingly small, it is easy to show that $h_{\rm max}= 9h_0$
\cite{walker}!  While this theoretical limit cannot be achieved, the dime can
easily hit the ceiling of a normal classroom when the superball is dropped
from chest level.

Another demonstration of this genre is to carefully place a baseball
(hardball) of mass $m$ on top of a basketball of mass $M$ and then drop the
two together.  Now the question for the class is: how high does the
basketball rise after collision with the floor?  It turns out that for
$m=M/3$ (which is close to the actual mass ratio for a baseball and a
basketball) and again for perfectly elastic collisions, the basketball hits
floor and stays there!  However, warn the class to beware of the rapidly
moving baseball!  In the theoretically ideal situation, it rises to a height
$h_{\rm max}= 4h_0$.  Working out these two examples is left as exercises for
the reader.

The extension of this simple two-particle system to arbitrary mass ratios
presents many interesting and unexpected challenges.  In what follows, we
generally ignore gravity because it plays a negligible role during the
collisions.  If the upper mass is much larger than the lower mass, then for
two separated masses that approach an elastic barrier, there will be a large
number of collisions before the two masses diverge and ultimately recede from
the barrier.  This dynamics can, in principle, be analyzed by applying
momentum conservation to map out the particle trajectories.  This approach is
tedious, however, and does not provide physical insight (see Appendix).

The goal of this article is to present a simple connection between the motion
of few-particle elastically colliding systems in one dimension and a
corresponding billiard system.  For two particles and an elastic barrier the
corresponding billiard ball moves in a two-dimensional wedge-shaped billiard
table with elastic and specular reflection each time the ball hits the
boundary of the table \cite{GZ,KT,T,G}.  Specular means that the angle of
incidence equals the angle of reflection.  This description can be greatly
simplified by recognizing that specular reflection at a boundary is
geometrically identical to passing straight through the boundary, where on
the other side of the boundary there is an identical image of the wedge.  By
repeating this construction, the end result is that billiard motion in the
wedge is equivalent to a straight trajectory in a plane that is ``tiled'' by
a fan of wedges.  By this equivalence, it is easy to completely solve the
collision history of the original two-particle and barrier system.

We then extend this approach to treat three elastically-colliding particles
of arbitrary masses on an infinite one-dimensional line.  Here we ask the
question: how many collisions occur when two cannonballs are approaching,
with an intervening elastic ping-pong ball that is rattling between them
\cite{S}?  This system can again be mapped onto the motion of a billiard ball
in an infinite wedge whose opening angle depends on the three masses.
Finally, we discuss the motion of three particles on a finite ring
\cite{GM,CA}.  This system can be mapped onto the motion of a billiard ball
on a triangular table.  Through this connection, we can gain many useful
insights about the collisional properties of the three particle on the ring.

These exactly soluble few-body systems naturally open new issues.  For
example, what happens when the number of particles becomes large?  In one
dimension, the momentum distribution of a polydisperse system of elastic
particles converges to a finite-$N$ version of the Gaussian distribution
\cite{RBF}.  However, transport properties appear to be anomalous.  In
particular, there is a lingering controversy about the nature of heat flow
through such a system \cite{heat} and the nature of the thermodynamic limit
of this system is not yet fully understood.

On a broader scope, one may naturally inquire about the roles of inelastic
collisions and the spatial dimension on the dynamics.  This is a natural
entry to the burgeoning field of granular media \cite{granular}.  While this
area is beyond the scope of this work, it is worth mentioning a few relevant
topics.  A particularly intriguing feature of inelastic systems is the
phenomenon of ``inelastic collapse'', where clumps of particles with
negligible relative motion form.  This collapse occurs when the number of
particles is sufficiently large or when collisions are sufficiently inelastic
\cite{MY,CGM,H}.  Some of the methods described here may be useful to
understand these systems.  The inclusion of other natural parameters also
leads to a wealth of new effects.  For example, a variety of collisional
transitions occur when inelastic particles are pushed by a massive wall
\cite{BM}, while the presence of gravity in an elastic system of two
particles and a wall leads to both quasi-periodic and chaotic behavior
\cite{WGC}.  Again, a billiard-theoretic perspective may provide helpful
insights into these systems.

In the next section, we discuss the dime on the superball and the baseball on
the basketball by elementary means.  Then in Sec.~III, we show how to the map
these systems onto the motion of a billiard ball in a wedge domain.  This
same approach is used to show the equivalence of three particles on an
infinite line to a billiard ball in an infinite wedge in Sec.~IV, and the
equivalence of three particles on a ring to a triangular billiard in Sec.~V.
A brief discussion is given in Sec.~VI.

\section{Dime on Superball \& Baseball on Basketball}

\begin{figure}[ht] 
 \vspace*{0.cm}
\includegraphics*[width=0.44\textwidth]{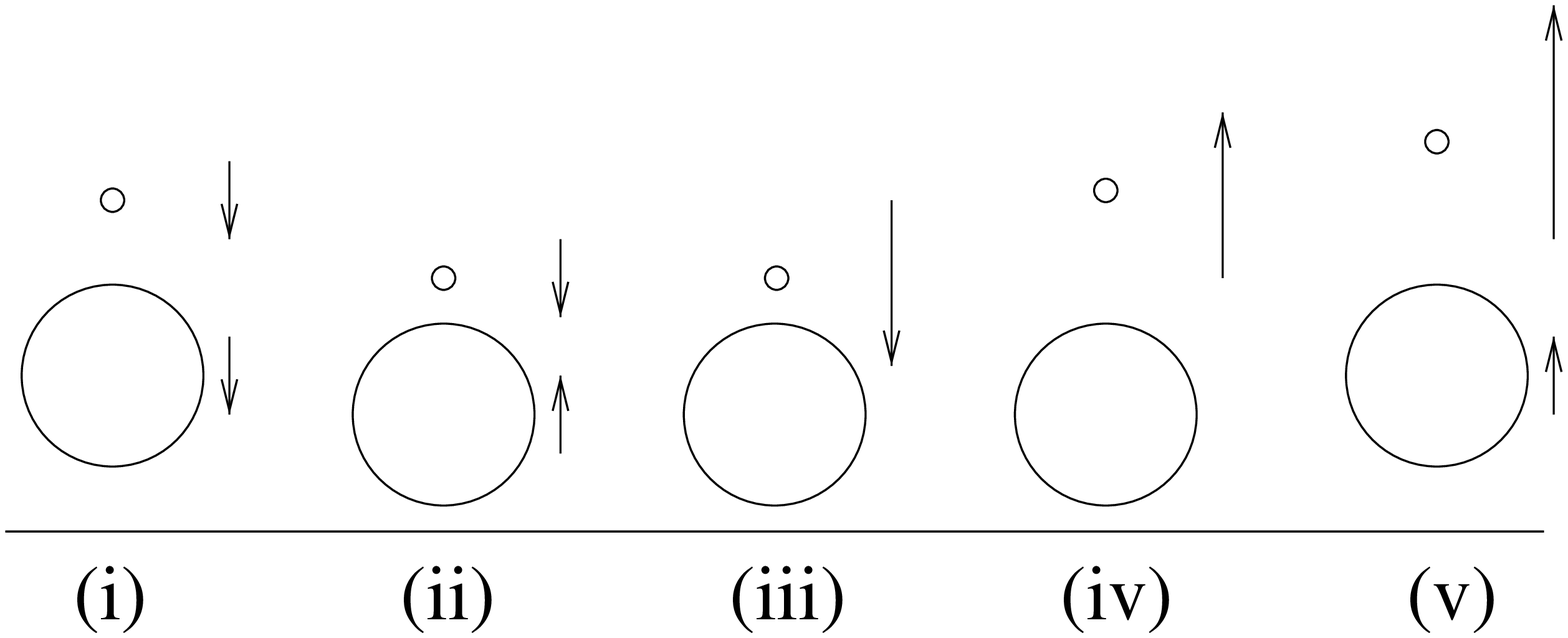}
\caption{Collision sequence for the dime-superball system.  The arrows 
  (drawn to scale) denote the velocities at each collision stage.
\label{system}}
\end{figure}

To aid in the analysis of the dime on the superball, it is helpful to imagine
that the two particles are separated.  Again, gravity is neglected throughout
the collisions; its only role is to give the final height of the dime
in terms of its velocity immediately after the last collision.
Fig.~\ref{system} shows the velocities of the dime and superball under the
assumption that the mass $m$ of the dime is negligible compared to that of
the superball ($M$), and that all collisions are perfectly elastic.  The
following collision sequence occurs:

\begin{itemize}
  
\item[(i)] The dime and the superball both approach the ground with velocity
  $-v$.
  
\item[(ii)] The superball hits the ground and reverses direction so that its
  velocity is $+v$.
  
\item[(iii)] For $m/M\to 0$, the center-of-mass coincides with the center of
  the superball.  In this reference frame, the dime approaches the superball
  with velocity $-2v$.
  
\item[(iv)] After the dime-superball collision in the center-of-mass frame,
  the dime moves with velocity $+2v$, while the superball remains at rest.
  
\item[(v)] Returning to the original lab frame, the superball moves with
  velocity $+v$, while the dime moves at velocity $+3v$.  This velocity of
  $+3v$ leads to the dime rising to a final height that is nine times that of
  the superball in the presence of gravity.

\end{itemize}

For the baseball on top of the basketball, the same analysis now gives the
following collision sequence (again assuming perfectly elastic collisions):

\begin{itemize}
\item[(i)-(ii)] The basketball hits the ground with velocity $-v$ and
  reverses direction so that its velocity is $+v$.
  
\item[(iii)] For $m/M=1/3$, the center-of-mass has velocity $+v/2$.  In the
  center-of-mass reference frame, the baseball has velocity $-3v/2$, while
  the basketball has velocity $+v/2$.
  
\item[(iv)] After the collision between the two balls in the center-of-mass
  frame, their velocities are reversed.
  
\item[(v)] In the original lab frame, the baseball has velocity $+2v$, while
  the basketball is at rest.

\end{itemize}

In both these cases, there are just two collisions -- an initial collision of
the lower ball with the floor and a second collision between the two balls.
Subsequently, the upper ball moves faster than the lower ball and there would
be no more collisions in the absence of gravity.  However, if the upper ball
is heavier than the lower ball, then there will be many collisions before the
two balls recede from the floor and from each other.  How many collisions
occur in total for this system?  What are the details of the collision
sequence?  These questions should be simple to answer, since only energy and
momentum conservation are involved.  However, when the upper ball is much
heavier than the lower ball the number of collisions is large and a direct
solution is tedious.  As we discuss in the next section, there is an elegant
mapping of this collision problem to an equivalent billiard system that
provides an extraordinary simplification.

\section{Billiard Mapping}

We now map the problem of two colliding particles and an elastic barrier into
an equivalent billiard system.  From this approach, the entire particle
collision history can be inferred in a simple geometric manner.  To be
general, suppose now that the particles have masses $m_1$ and $m_2$ and are
located, respectively, at $x_1$ and $x_2$, with $x_1< x_2$ (and $x_1,
x_2>0$).  The trajectories of the two particles in one dimension are
equivalent to the trajectory $(x_1(t),x_2(t))$ of an effective billiard ball
in the two-dimensional domain defined by $x_1, x_2>0$ and $x_1<x_2$.  The
billiard ball hitting the boundary $x_1=0$ corresponds to a collision between
the lower particle and the floor, while the ball hitting the boundary
$x_1=x_2$ corresponds to a collision between the two particles.

\begin{figure}[ht] 
 \vspace*{0.cm}
\includegraphics*[width=0.2\textwidth]{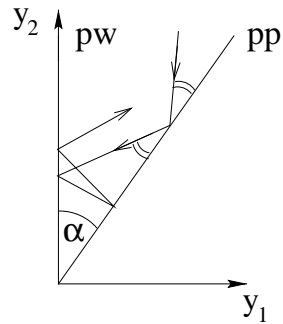}
\caption{Allowed wedge in $y_i$ coordinates.  A sample
  billiard ball trajectory is shown.  Hitting the $y_2$ axis corresponds to a
  particle-wall collision (denoted by $pw$), while hitting the line $y_2=
  y_1\sqrt{{m_2}/{m_1}}$ corresponds to a particle-particle collision
  (denoted by $pp$).
\label{wedge}}
\end{figure}

Now define the following ``billiard rescaling'' \cite{GZ,KT,T,G}:
\begin{eqnarray*}
\label{tform}
y_i=x_i\,\sqrt{m_i}\qquad w_i=v_i\; \sqrt{m_i},
\end{eqnarray*}
for $i=1,2$.  In these coordinates, the constraint $x_1 <x_2$ becomes
\begin{eqnarray*}
\label{constraint}
y_2> \sqrt{\frac{m_2}{m_1}} \;y_1.
\end{eqnarray*}
Thus the allowed region is now a wedge-shaped domain (see Fig.~\ref{wedge})
with opening angle
\begin{equation}
\label{alpha-2p}
\alpha\equiv \tan^{-1}\sqrt{\frac{m_1}{m_2}}.
\end{equation}

The crucial feature of this rescaling is that it ensures that all collisions
of the billiard ball with boundary of the domain are specular.  To
demonstrate this point, we take the energy and momentum conservation
statements,
\begin{eqnarray}
\label{cons}
\frac{1}{2}\, m_1v_1^2+\frac{1}{2}\, m_2v_2^2 &=& 
\frac{1}{2}\, m_1{v_1'\,^2}+\frac{1}{2}\, m_2{v_2'\,^2}\nonumber \\
m_1v_1+m_2v_2 &=& m_1v_1'+m_2v_2',
\end{eqnarray}
where the prime denotes a particle velocity after a collision, and rewrite
these conservation laws in rescaled coordinates to give
\begin{eqnarray}
\label{cons-tr}
w_1^2+w_2^2 &=& {w_1'\,^2}+{w_2'\,^2}\nonumber \\ 
\sqrt{m_1}\,w_1+\sqrt{m_2}w\,_2 &=& \sqrt{m_1}\,w_1'+\sqrt{m_2}\,w_2'.
\end{eqnarray}

The first of these equations states that the speed of the billiard ball is
unchanged by a collision.  The second equation can be rewritten as
$(\sqrt{m_1},\sqrt{m_2})\cdot (w_1,w_2)$ remains constant in a collision.
Since the vector $(\sqrt{m_1},\sqrt{m_2})$ is tangent to the constraint line
$y_2= y_1\sqrt{{m_2}/{m_1}}$, the projection of the rescaled velocity onto
this line is constant in a particle-particle collision.  It is also
intuitively clear that in a particle-wall collision the rescaled velocity is
also preserved.  As a result, the collision sequence of two
elastically-colliding particles and an elastic barrier in one dimension is
completely equivalent to the trajectory of a billiard ball in a
two-dimensional wedge of opening angle $\alpha$ in which each collision with
the boundary is specular.

\begin{figure}[ht] 
 \vspace*{0.cm}
\includegraphics*[width=0.3\textwidth]{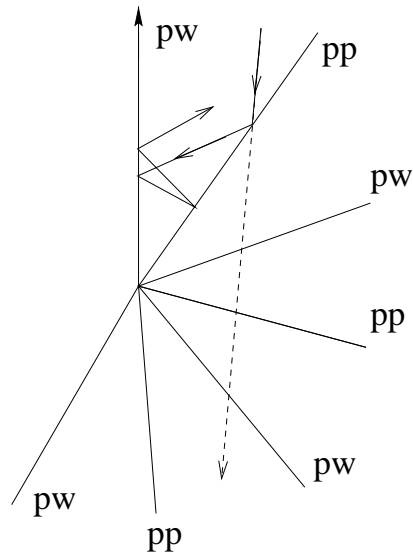}
\caption{Periodic extension of the allowed wedge.  The trajectory in the
  original wedge is equivalent to the straight trajectory shown (dashed).
\label{wedge-periodic}}
\end{figure}

A more dramatic simplification arises in the $y_i$ coordinates by recognizing
that since each reflection is specular, the trajectory in the wedge is the
same as a straight trajectory in the periodic extension of the wedge
(Fig.~\ref{wedge-periodic}).  Each collision is alternately a
particle-particle or a particle-wall, so that the identity of each barrier
alternates between $pp$ and $pw$.  From this description, we immediately
deduce that the collision sequence of the two-particle system ends when the
trajectory of the billiard ball no longer crosses any wedge boundary.  As
shown in Fig.~\ref{wedge-periodic}, when the original trajectory is extended
in the manner, it will ultimately pass through six wedges.  Thus five
collisions (particle-wall and particle-particle) occur in total.

The maximum number of wedges that can be packed in the half plane is
$\pi/\alpha$.  A straight trajectory of the billiard ball typically passes
through all these wedges.  This is therefore the maximum number of collisions
$N_{\rm max}$ possible in the two-particle system.  In the limit where
$m_1\ll m_2$, we thereby find (using Eq.~(\ref{alpha-2p})
\begin{equation}
\label{nmax}
N_{\rm max}= \simeq \pi\sqrt{\frac{m_2}{m_1}}.
\end{equation}
Thus the total number of collisions in the original particle system emerges
from extremely simple geometric considerations of the equivalent billiard.

From Fig.~\ref{wedge-periodic}, the incidence angle of the billiard ball at
each boundary increases by a factor $\alpha$ after each collision.
Furthermore, using the constancy of the rescaled velocity $w$, one can also
deduce the particle velocities at every collision stage.  We now illustrate
this approach by reconsidering our initial examples from this
billiard-theoretic perspective.

\section{Dime on Superball \& Baseball on Basketball: A Second Look}

\begin{figure}[ht] 
  \vspace*{0.cm} \includegraphics*[width=0.44\textwidth]{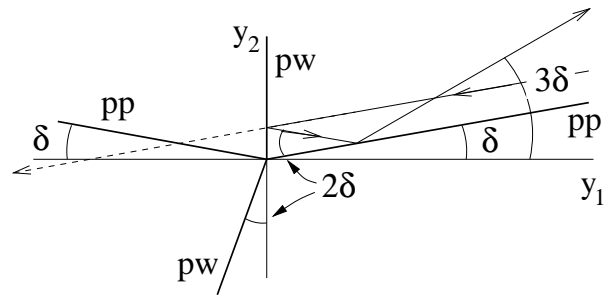}
\caption{Allowed wedge in $y_i$ coordinates for the
  dime and superball.  A trajectory corresponding to the dime and the
  superball approaching the wall at the same velocity is shown.  The dashed
  line shows the trajectory in the periodic extension of the wedges.
\label{ds}}
\end{figure}

For the dime on the superball, the opening angle of the wedge has the
limiting behavior $\alpha=\frac{\pi}{2}-\delta$, with $\delta\approx
\sqrt{{m_2}/{m_1}}$ as $m_2/m_1\to 0$.  In Fig.~\ref{ds}, the trajectory
of the corresponding billiard ball is shown in the $y_1$-$y_2$ coordinate
system.  Because the dime and the superball have the same initial velocities,
the incoming trajectory in the wedge is parallel to the initial $pp$
boundary.  The distance to the $pp$ boundary is proportional to the initial
separation of the dime and the superball.  (If, initially, $x_2=x_1+\epsilon$,
then $y_2= y_1\sqrt{{m_2}/{m_1}}+\sqrt{m_2}\,\epsilon$.)

After the superball collides with the wall, the billiard trajectory is
incident on the $pp$ boundary with inclination angle $2\delta$
(Fig.~\ref{ds}).  After specular reflection from this boundary, the final
outgoing trajectory is then inclined at an angle $3\delta$ with respect to
the horizontal.  This inclination angle means that the final velocity of the
dime is three times that of the superball.  Thus in the presence of gravity,
an ideal dime will rise to nine times its initial height.

This same result can be obtained even more simply by drawing a straight
trajectory through the periodic extension of the wedges.  In this case, the
final trajectory is inclined at an angle of $\frac{\pi}{2}-3\delta$ with
respect to the last periodically extended $pw$ boundary.  This construction
again implies that the outgoing trajectory is inclined at an angle of
$3\delta$ with respect to the initial $pp$ boundary.

\begin{figure}[ht] 
  \vspace*{0.cm} \includegraphics*[width=0.36\textwidth]{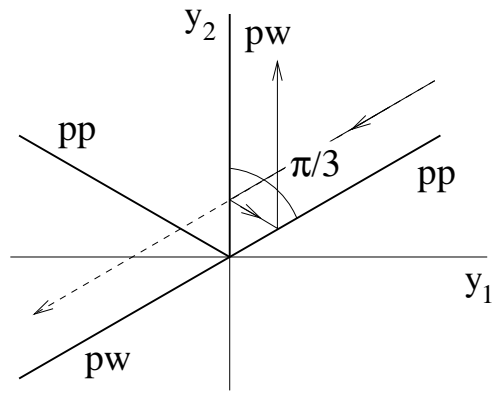}
\caption{Allowed wedge in $y_i$ coordinates for the
  baseball-basketball system with $m_1=3m_2$.  The dashed line again shows
  the billiard trajectory in the periodic extension of the wedges.
\label{bb}}
\end{figure}

For a basketball of mass $m_1=3m$ and a baseball of mass $m_2=m$, the opening
angle of the wedge is now $\alpha=60^\circ$ (Fig.~\ref{bb}).  Again, there
are two collisions in total and by simple geometry it easily follows that the
final outgoing billiard trajectory is vertical, {\it i.e.}, $v_1'=0$ and
$v_2'>0$.  We can obtain the final speed $v_2'$ by exploiting the constancy
of the rescaled speed.  Initially, $\sqrt{w_1^2+w_2^2}=
\sqrt{m_1v^2+m_2v^2}=2\sqrt{m}\,v$, where $v$ is the initial velocity.  In
the final state, the rescaled speed is $w_2'=\sqrt{m}\,v_2'$.  Therefore
$v_1'=0$ and $v_2'=2v$.

As a byproduct of the billiards approach, notice that as soon as $m_1/m_2<3$,
the total angle of three wedges is less that $180^\circ$ and there
necessarily must be one more $pw$ collision.  Whenever the final trajectory
is tangent to either a $pp$ or a $pw$ line, a critical point is defined where
the total number of collisions changes by one.  As $m_1$ continues to
decrease, a sequence of transitions arises.  Each transition occurs when the
wedge angle decreases below $\pi/n$, with $n$ an integer.  At this point the
total number collisions increases from $n-1$ to $n$.  We therefore find that
three collisions first occur when $m_1<3m_2$, four when $m_1<m_2$, five when
$m_1<0.5278 m_2$, six when $m_1<m_2/3$, {\it etc.}

\section{Three Particles on an Infinite Line}

\begin{figure}[ht] 
  \vspace*{0.cm} \includegraphics*[width=0.4\textwidth]{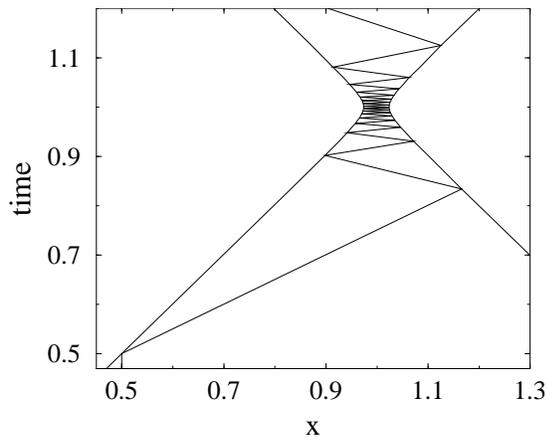}
\caption{Space-time diagram of the typical evolution of two 
  cannonballs (heavy lines) approaching an initially stationary ping-pong
  ball (light line).  The cannonballs each have mass $M=1$ and initial
  conditions $(x_1(0),v_1(0))=(0,1)$ and $(x_3(0),v_3(0))=(2,-1)$.  A
  ping-pong ball of mass $m=0.005$ is initially at $x_2(0)=1/2$.  There are
  31 collisions in total before the three particles recede.  The first 30
  collisions are shown.
\label{cpc}}
\end{figure}

The billiards approach gives an extremely simple way to solve a classic
elastic collision problem that was apparently first posed by Sinai \cite{S}.
Consider a three-particle system on an infinite line that consists of two
approaching cannonballs, each of mass $M$.  Between them (and
non-symmetrically located) lies a ping-pong ball of mass $m\ll M$.  Due to
the collisions between the cannonballs and the intervening ping-pong ball,
the latter rattles back and forth with a rapidly increasing speed until its
momentum is sufficient to drive the cannonballs apart (Fig.~\ref{cpc}).  In
the final state, the three particles are receding from each other.  How many
collisions occur before this final state is reached?

Using energy and momentum conservation, we can determine the state of the
system after each collision and thereby find the number of collisions before
the three particles mutually recede.  However, this approach is complicated
and provides minimal physical insight (see Appendix and also
\cite{J,M,CGM,H}).  We now present a much simpler solution by mapping the
original three-particle system onto a billiard in an appropriately-defined
domain.

We denote the coordinates of the particles as $x_1$, $x_2$, and $x_3$, with
$x_1< x_2< x_3$.  This order constraint between the particles again
translates to a geometrical constraint on the accessible region for the
billiard ball in the three-dimensional $x_i$ space.  Similarly, the
trajectories of the particles on the line translate to the trajectory
$(x_1(t),x_2(t),x_3(t))$ of a billiard ball in the allowed region.

\begin{figure}[ht] 
 \vspace*{0.cm}
\includegraphics*[width=0.4\textwidth]{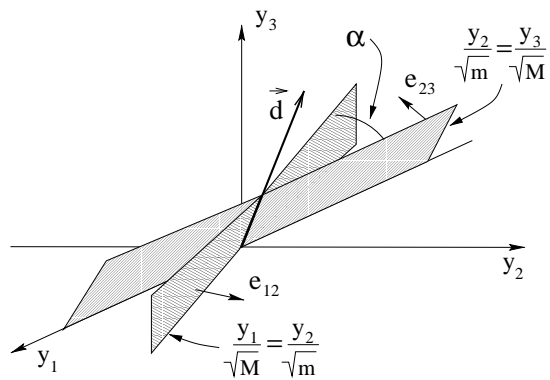}
\caption{Allowed wedge in the $y_i$ coordinate system for a system of
  two cannonballs and an intervening ping-pong ball on an infinite line.  The
  wedge is the region where the vectors $e_{12}$ and $e_{23}$ point towards.
\label{planes}}
\end{figure}

As in the previous examples, we introduce the rescaled coordinates
$y_i=x_i\sqrt{m_i}$.  These coordinates then satisfy the constraints
\begin{eqnarray*}
\frac{y_1}{\sqrt{M}} < \frac{y_2}{\sqrt{m}}\,,\qquad 
\frac{y_2}{\sqrt{m}} < \frac{y_3}{\sqrt{M}}\,.
\end{eqnarray*}
(The generalization to arbitrary masses is straightforward and is made in the
next section.)~ In $y_i$ space, the constraints correspond, respectively, to
the effective billiard ball being confined to the half-space to the right of
the plane ${y_1}/{\sqrt{M}}= {y_2}/{\sqrt{m}}$ and to the half-space to the
left of the plane ${y_2}/{\sqrt{m}}= {y_3}/{\sqrt{M}}$, as illustrated in
Fig.~\ref{planes}.  This defines the allowed region as an infinite wedge of
opening angle $\alpha$

The use of rescaled coordinates ensures that all collisions between the
effective billiard particle and these constraint planes are specular.
Further, momentum conservation gives 
\begin{eqnarray*}
Mv_1+mv_2+Mv_3=\sqrt{M}\, w_1+\sqrt{m}\, w_2+\sqrt{M}\, w_3= 0,
\end{eqnarray*}
where, without loss of generality, we take the total momentum to be zero.  In
this zero momentrum reference frame, the trajectory of the billiard ball is
always perpendicular to the diagonal, $\vec d=(\sqrt{M},\sqrt{m},\sqrt{M})$.
Thus we may reduce the three-dimensional billiard to a two-dimensional system
in the plane perpendicular to $\vec d$.

To complete this picture, we need to find the wedge angle $\alpha$.  The
normals to the two constraint planes are $\vec e_{12}=
(-\frac{1}{\sqrt{M}},\frac{1}{\sqrt{m}},0)$ and $\vec e_{23}=
(0,-\frac{1}{\sqrt{m}},\frac{1}{\sqrt{M}})$.  Consequently, the angle between
these planes is given by
\begin{eqnarray}
\label{alpha}
\alpha &=&\cos^{-1}\left(-\frac{\vec e_{12}\cdot\vec e_{23}}{|\vec e_{12}|\;
    |\vec e_{23}|}\right)\nonumber \\
& =&\cos^{-1}\left(\frac{1}{{\displaystyle1+\frac{m}{M}}}\right).
\end{eqnarray}
In the limit $m/M\to 0$, this gives $\alpha\approx\sqrt{2m/M}$.  Finally, the
maximum number of possible collisions is determined by the number of wedges
that fit into the half plane.  This gives
\begin{equation}
\label{N}
N_{\rm max}=\frac{\pi}{\alpha}\approx \pi\sqrt{\frac{M}{2m}}.
\end{equation}
For $m/M\to 0$ the opening angle of the wedge goes to zero and
correspondingly, the number of collisions diverges.

\section{Three Particles on a Ring and the Triangular Billiard}

\begin{figure}[ht] 
 \vspace*{0.cm}
\includegraphics*[width=0.4\textwidth]{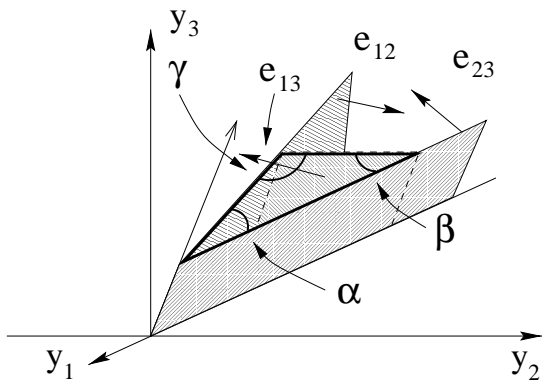}
\caption{Allowed region in the $y_i$ coordinates for three
  particles of arbitrary masses on a ring of circumference $L$.  The
  triangular billiard with angles $\alpha$, $\beta$ and $\gamma$ is defined
  by the thick solid lines.
\label{triangle}}
\end{figure}

Finally, let us consider three elastically colliding particles of arbitrary
masses $m_1$, $m_2$, and $m_3$ on a finite ring of length $L$ \cite{GM,CA}.
If we make an imaginary cut in the ring between particles 1 and 3, then we
can write the order constraints of the three particles as
\begin{eqnarray*}
x_1< x_2, \qquad x_2< x_3, \qquad x_3< x_1+L
\end{eqnarray*}
As usual, we employ the rescaled coordinates $y_i=x_i\sqrt{m_i}$ to ensure
that all collisions of the billiard ball with the domain boundaries in the
$y_i$ coordinates are specular.  In these coordinates, the first two
constraints again confine the particle to be between the planes defined by
the normal vectors $\vec e_{12}=
(-\frac{1}{\sqrt{m_1}},\frac{1}{\sqrt{m_2}},0)$ and $\vec e_{23}=
(0,-\frac{1}{\sqrt{m_2}},\frac{1}{\sqrt{m_3}})$.  Without the offset of $L$,
the constraint $ x_3< x_1+L$ corresponds to a plane that slices the
$y_1$-$y_3$ plane and passes through the origin.  The offset of $L$ means
that we must translate this plane by a distance $L\sqrt{m_3}$ along $y_3$.
The fact that $x_3$ is the lesser coordinate also means that the billiard
ball is confined to the near side of this constraint plane.  Thus the
billiard ball must remain within a triangular bar whose outlines are shown in
Fig.~\ref{triangle}.

If the total momentum of the system is zero, then
$(\sqrt{m_1},\sqrt{m_2},\sqrt{m_3})\cdot (w_1,w_2,w_3)=0$ and the trajectory
of the billiard ball remains within a triangle perpendicular to the long axis
of the bar, with angles $\alpha$, $\beta$, and $\gamma$.  We compute these
angles by the same approach given in Eq.~(\ref{alpha}).  Thus, for example,
\begin{eqnarray}
\label{alpha-gen}
\alpha &=&\cos^{-1}\left(-\frac{\vec e_{12}\cdot\vec e_{23}}{|\vec e_{12}|\;
    |\vec e_{23}|}\right)\nonumber \\ \nonumber\\
&=& \cos^{-1}\left(\sqrt{\frac{m_2 m_3}{(m_1+m_2)(m_2+m_3)}}\right).
\end{eqnarray}
The angles $\beta$ and $\gamma$ can be obtained by cyclic permutations of
this formula.

Therefore the elastic collisions of three particles on a finite ring can be
mapped onto the motion of a billiard ball within a triangular billiard table.
One can then exploit the wealth of knowledge about triangular billiards
\cite{T,G,GM} to infer basic collisional properties of the three-particle
system.  For example, periodic or ergodic behavior of the billiard translates
to periodic or non-periodic behavior in the three-particle collision
sequence.

\section{Discussion}

We have shown how to recast the elastic collisions of point particles in one
dimension into the motion of a billiard ball that moves at constant speed in
a confined region of a higher-dimensional space.  A crucial step in this
reformulation is to introduce the rescaled coordinates $y_i=x_i\sqrt{m_i}$.
This rescaling ensures that all collisions of the billiard ball with the
boundaries of its accessible region are specular.  For the examples of two
particles on a semi-infinite line and a reflecting wall and for three
particles on an infinite line, the allowed region for the billiard ball is an
infinite wedge.  For three particles on a ring, the allowed region is a
triangular billiard.  The shape of the associated wedge or triangle is
readily calculable in terms of the particle masses in the original system.

When these masses are widely disparate, the opening angle of the wedge or one
angle in the triangle becomes small.  If a billiard ball enters such an acute
corner, a large number of bounces occurs before the ball recedes from this
corner.  These frequent bounces are completely equivalent to a straight line
trajectory passing close to the tips of a large number of periodic extensions
of the wedge over a short distance.  In the original system, either picture
corresponds to a large number of collisions between neighboring particles.

The mapping onto a billiard system can, in principle, be generalized to an
arbitrary number of particles $N$.  The spatial dimension of the accessible
region in the corresponding billiard is now $N-1$-dimensional.  While less is
known about such high dimensional billiards, this mapping provides a useful
perspective to deal with the elastic collisions of many particles in one
dimension.

Finally, it is worth mentioning that a similar wedge mapping has been applied
to determine the probability that three diffusing particles on the line obey
various constraints on their relative positions \cite{redner}.  In both the
collisional and diffusive systems, the order constraint leads to nearly
identical wedge constructions, and these provide elegant solutions to the
original respective problems.

\section{Acknowledgments}

I thank T. Antal, S. Glashow, P. Hurtado, and P. Krapivsky for pleasant
discussions, and P. Hurtado for helpful manuscript comments.  I am grateful
to NSF grant DMR0227670 for partial support of this work.

\appendix

\section{Ping-Pong Ball Between Two Cannonballs: Direct Solution}

\begin{figure}[ht] 
 \vspace*{0.cm}
\includegraphics*[width=0.3\textwidth]{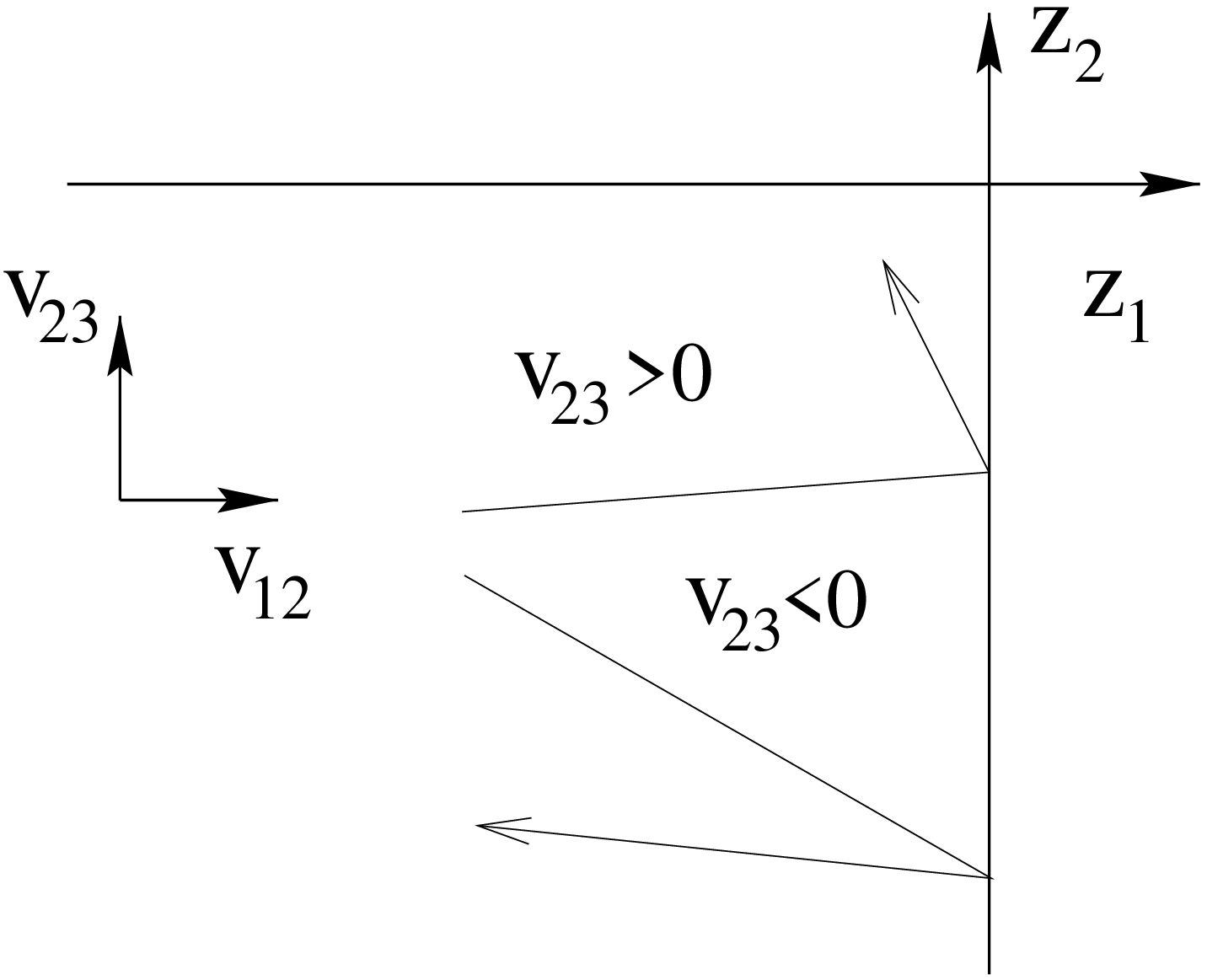}
\caption{Allowed region in $z_1$-$z_2$ coordinates for three
  particles of arbitrary masses on the line.  Typical non-specular collisions
  of the billiard ball with the boundary are shown for 12 collisions when
  $v_{23}>0$ and $v_{23}<0$ (see Eq.~(\ref{c12})).  A similar picture arises
  for 23 collisions.
\label{quadrant}}
\end{figure}

We consider three particles with masses $m_i$, positions $x_i$ and velocities
$v_i$ for $i=1,2,3$.  Define the relative coordinates $z_i=x_i-x_{i+1}<0$ and
the relative velocities $v_{i,i+1}=x_i-x_{i+1}$.  From elementary mechanics,
these relative velocities transform as follows (see Fig.~\ref{quadrant}):
\begin{eqnarray}
\label{c12}
12\ {\rm collision}:\quad \begin{array}{ll}
 v_{12}'=-v_{12} \\ v_{23}'=v_{23}+\lambda_{12} v_{12},
\end{array}
\end{eqnarray}
\begin{eqnarray}
\label{c23}
23\ {\rm collision:}\quad \begin{array}{ll}
v_{12}'=v_{12}+\lambda_{23} v_{23} \\ v_{23}'=-v_{23},
\end{array}
\end{eqnarray}
with $\lambda_{12} =2m_1/(m_1+m_2)$ and $\lambda_{23} =2m_3/(m_2+m_3)$.  Once
again, we can view the particle collisions as equivalent to the motion of a
billiard ball in the third quadrant of the $z$-plane, but with non-specular
reflections at each boundary.

It is convenient to characterize a trajectory by its polar angle
$\tan\theta=v_{23}/v_{12}$.  Then the above collision rules can be written as 
\begin{eqnarray}
\label{basic}
&12& {\rm collision}:  \tan\theta_n=-\lambda_{12}-\tan\theta_{n-1},\nonumber \\
&23& {\rm collision}:  \cot\theta_{n-1}=-\lambda_{23}^{-1}-\cot\theta_{n-2},
\end{eqnarray}
where $\theta_n$ is the angle after the $n^{\rm th}$ collision with the
boundary.  In writing these recursions, we use the fact that the 12 and 23
collisions alternate.  Initially, the billiard ball is heading toward the
corner, but eventually it ``escapes'' by having a trajectory with polar angle
in the range $(\pi,3\pi/2)$.  This condition means that the three particles
are all receding from each other.  For a 12 collision, the incidence angle is
in the range $-\frac{\pi}{2}<\theta<\frac{\pi}{2}$ while the outgoing angle
is in the range $\frac{\pi}{2}<\theta<\frac{3\pi}{2}$.  In this case, escape
means that $\tan\theta>0$.  Similarly, for a 23 collision, the incidence
angle is in the range $0<\theta<\pi$ while the outgoing angle is in the range
$\pi<\theta<2\pi$.  For this case, escape means that $\cot\theta>0$.

To solve Eqs.~(\ref{basic}), we first write $\theta_n$ in terms of
$\theta_{n-2}$:
\begin{equation}
\label{n1}
\tan\theta_{n}=-\lambda_{12}+
\frac{1}{{\displaystyle\frac{1}{\lambda_{23}}}+{\displaystyle \frac{1}{\tan\theta_{n-2}}}}\,.
\end{equation}
Next, let $t_n=\tan\theta_n/\sqrt{\lambda_{12}\lambda_{23}}$.  This
simplifies Eq.~(\ref{n1}) to 
\begin{equation}
\label{simple}
t_{n}=-\mu+{\displaystyle \frac{1}{{\displaystyle \mu+\frac{1}{t_{n-2}}}}}\,,
\end{equation}
where $\mu=\sqrt{\lambda_{12}/\lambda_{23}}$.  This equation can be written
even more simply as
\begin{equation}
\label{simpler}
t_{n}=-\mu-\frac{1}{t_{n-1}}\,.
\end{equation}

To solve this recursion formula, we define $t_n\equiv g_n/h_n$ and find that
Eq.~(\ref{simpler}) is equivalent to the two first-order recursion relations
$h_n=g_{n-1}$ and $g_n=-\mu g_{n-1}-h_{n-1}$.  This, in turn, is equivalent
to the second-order recursion
\begin{equation}
\label{g}
g_{n}=-\mu g_{n-1}-g_{n-2}.
\end{equation}
The general solution is $g_n=A_+\alpha_+^n+A_-\alpha_-^n$, where $A_\pm$ are
constants and $\alpha_\pm=(-\mu\pm\sqrt{\mu^2-4})/2$.  To complete the
solution, we need an initial condition.  For simplicity, we start with a
billiard with incidence angle $\theta=0$ that has undergone a single 12
collision.  This situation corresponds to the initial condition $t_1=-\mu$.
Imposing this condition, and performing some simple algebra, we find
\begin{equation}
\label{t-alpha}
t_{n}=\frac{\alpha_+^{n+1}-\alpha_-^{n+1}}{\alpha_+^{n}-\alpha_-^{n}}.
\end{equation}
It is more convenient to write this in complex form by defining
$\alpha_{\pm}=A\,e^{\pm i\phi}$.  This then leads to
\begin{equation}
\label{t-final}
t_{n}=\frac{\sin(n+1)\phi}{\sin n\phi},
\end{equation}
where $\phi=\tan^{-1}\sqrt{(4-\mu^2)/\mu^2}$.

The particles are all receding when $t_n$ first becomes negative.  The
maximum number of collisions until this occurs is thus given by the condition
$t_n=0$; this gives $(n+1)\phi=\pi$, or $n\approx \pi/\phi$.  In the limit
$m_2/m_1\equiv\epsilon_1\to 0$ and $m_2/m_3\equiv\epsilon_3\to 0$, we have
\begin{eqnarray*}
\mu^2&=&\frac{4}{(1+\epsilon_1)(1+\epsilon_3)}\\
 &=& 4(1-\epsilon_1-\epsilon_3).
\end{eqnarray*}
Finally
$\phi=\tan^{-1}\sqrt{(4-\mu^2)/\mu^2}\approx\sqrt{\epsilon_1+\epsilon_3}$.
This then gives
\begin{equation}
N_{\rm max}\approx \frac{\pi}{\sqrt{\epsilon_1+\epsilon_3}}
=\pi\sqrt{\frac{m_1m_3}{m_2(m_1+m_3)}}.
\end{equation}
In the special case of $m_1=m_3=M$ and $m_2=m$, this expression reduces to
Eq.~(\ref{N}).

\end{document}